# Photoluminescence Dynamics of CdSe QD/polymer Langmuir-Blodgett Thin Films: Morphology Effects

*Beatriz Martín-García,[1] Pedro M. R. Paulo,[2,*] Sílvia M. B. Costa[2] and M. Mercedes Velázquez[1]*

[1] Departamento de Química Física, Facultad de Ciencias Químicas, Universidad de Salamanca, E-37008-Salamanca, Spain

[2] Centro de Química Estrutural, Instituto Superior Técnico, Universidade Técnica de Lisboa, Av. Rovisco Pais 1, 1049-001 Lisboa, Portugal. *E-mail: pedro.m.paulo@ist.utl.pt

*Abstract*

Thin films of colloidal semiconductor CdSe quantum-dots (QDs) and a styrene/maleic anhydride copolymer were prepared by the Langmuir-Blodgett technique and their photoluminescence was characterized by confocal fluorescence lifetime microscopy. In the films, the photoluminescence dynamics are strongly affected by excitation energy migration between close-packed quantum-dots and energy trapping by surface-defective QDs or small clusters of aggregated QDs. Polymer/QD films with more aggregated QD regions exhibit lower PL intensities and faster decays, which we attribute primarily to the increased ability of excitations to find trap sites among more close-packed QD regions. This was confirmed by comparing films from bilayer or co-spreading deposition, and varying the QD-to-polymer composition or the surface pressure at deposition. The more regular films, such as those obtained by bilayer deposition at high-pressure, display more surface emission intensity and decays with less accentuated curvature. Decay analysis was performed with a model that accounts for excitation energy migration and trapping in the films. In the framework of the model, the photoluminescence dynamics are related to film morphology through the density of energy traps. A lower amount of QD clustering in the films reflects on a lower density of energy traps and, thus, on more emissive QD films, as inferred here for bilayer films relatively to co-spreading films.

*Introduction*

Colloidal semiconductor CdSe quantum-dots (QDs) are widely used in optoelectronic devices such as LEDs, photodetectors, solar cells and nanosensors.[1-3] For the construction of nano-structured devices with semiconductor QDs, it is necessary to obtain large ordered structures with homogeneous and known photophysical properties.[4] Therefore, it is important to understand the influence of particle density, inter-dot distance and environmental properties on the photophysics of QD assemblies.

Some optoelectronic applications make use of QD properties by incorporating them into a polymeric matrix to build nanocomposites. In these cases, it is necessary to develop a controlled way to achieve good-quality QD/polymer hybrid films.[3,5] In order to optimize such hybrid systems, several authors have focused on the effect of film matrices on energy transfer processes between QDs [6,7] and on their photoluminescence intermittency.[8,9] An important issue concerning the properties of QDs deposited on solids is achieving control over the organization and assembly of nanoparticles at interfaces. In film organization, the interparticle distance is paramount for applications, such as light-harvesting in photovoltaic cells, because short interparticle distances improve the energy transfer in the QDs films.[10-12] On the other hand, the film morphology also plays an important role on the photophysical properties because surface defects or aggregation of dots may provide excitation energy traps that can efficiently reduce photoluminescence. Within several methodologies proposed, the Langmuir-Blodgett (LB) technique has demonstrated to be an excellent candidate to construct good quality QD thin films, because it offers the possibility of preparing reproducible films with control of the interparticle distance.[13,14]

In the current work, we are interested to study the effect of the morphology of QD 4

assemblies on the photoluminescence properties of QD films. Previously, we have prepared and characterized hybrid films of QDs and polymers derived of maleic anhydride and the Gemini surfactant 18-2-18 by the LB method.[15,16] Here, we study the photoluminescence properties of QD films prepared with the polymer poly(styrene-co-maleic anhydride) partial 2-buthoxy ethyl ester cumene terminated. We have chosen this polymer because it renders QD assemblies with different morphology using two different preparation methods, co-spreading and bilayer deposition. Moreover, styrene/maleic anhydride copolymers have shown potential application in optical waveguides, electron beam resists and photodiodes.[17,18]

The photoluminescence dynamics of the QD/polymer films were measured by confocal fluorescence lifetime microscopy. Time-resolved photoluminescence has the advantage of being a technique highly sensitive to surface and environmental changes, but this information is often concealed by the intrinsically complex excited-state dynamics of QDs.[19] It is common to use phenomenological fitting functions in decay analysis, e.g. stretched exponentials or statistical decay time distributions, which afford good fitting results, but are more questionable in terms of extracting physical insight from the fitted parameters. Instead, we have adopted an approximate model for decay analysis that considers excitation energy transport and trapping effects. The parameters obtained from the decay fits were used to rationalize and compare results of the different QD/polymer films and to relate their photoluminescence properties to the film morphology previously characterized.

*Experimental Section*

**Materials.** The TOPO-capped CdSe QDs were synthesized by the method proposed by Yu and Peng.[20] The QDs thus obtained are capped by X-type octylphosphonate derivatives, which remain attached to the CdSe QD surface after the purification and precipitation steps.[21-23] QDs were collected as powder by size-selective precipitation with acetone and dried under vacuum. The polymer poly(styrene-co-maleic anhydride) partial 2-buthoxy ethyl ester cumene terminated, PSMABEE hereafter, was purchased from Sigma-Aldrich and used as received. The PSMABEE ester:acid ratio 1:1 and the molecular weight $M_n$ = 2.5 kDa were provided by the manufacturer. Chloroform (PAI, filtered) used to prepare the spreading solutions was purchased from Sigma-Aldrich. Millipore Ultra pure water prepared using a combination of RiOs and Milli-Q systems from Millipore was used as subphase. The LB quartz substrates were supplied by TedPella (USA) and were cleaned before use by the RCA procedure.[24] First, the substrates were successively cleaned with acetone (PAI), ethanol (PAI) and MilliQ water. Next, in a glass beaker, Milli-Q water and ammonia (25% vol) (5:1 vol/vol) were mixed up to 60 mL and heated until 70ºC, and then 10 mL of hydrogen peroxide (30%w) was added. The substrates were submerged in the cleaning mixture and maintained during 15 min at 70ºC. Afterwards, the substrates were rinsed with abundant Milli-Q water and dried under a stream of nitrogen.

**Film preparation.** The samples were prepared by the Langmuir-Blodgett technique on a Langmuir standard-trough (KSV2000 System 2, Finland) placed in an anti-vibration table. Monolayers were transferred by symmetric barrier compression (5 mm min$^{-1}$) with the quartz substrate into the trough by vertically dipping it upwards. Spreading solution was deposited onto the water subphase with a Hamilton microsyringe with a precision of 1 μL. The surface pressure, π, was measured with a Pt-Wilhelmy plate connected to an electrobalance. The subphase temperature was maintained at (23.0 ± 0.1) ºC by flowing thermostated water through jackets at the bottom of the trough. Surface pressure-area isotherms for the pure components, CdSe QDs and PSMABEE, agree with those reported previously[25,26] and their mixtures were previously characterized.[15,16] Two different deposition methods were used: *i*) co-spreading deposition, the mixed QD/polymer monolayers are transferred from the air-water interface onto quartz, or *ii*) bilayer deposition, the QDs are transferred onto a polymer film already on the surface. To prepare the bilayer films, firstly a PSMABEE monolayer is LB transferred on quartz substrate, and dried under vacuum at least during 12 h, before the LB deposition of the QD monolayer.

In order to avoid spurious variations in the emission quantum yield of the QDs caused by solvent or dilution effects on the surface conditions of the QDs,[22,27-29] we have only used one batch of synthesized QDs. All the spreading solutions were prepared in chloroform with the QD powder obtained from a single extraction. The QD concentration in the spreading solutions was the same for all samples: bilayer and co-spreadings depositions, ca. $10^{-6}$M.

TEM images of the LB films deposited on Formvar®-carbon coated copper grids were taken with 80 kV TEM (ZEISS EM 902, Germany). The LB deposition onto copper grids was carried out at a speed up of 1 mm min$^{-1}$.

**Confocal fluorescence lifetime microscopy.** Surface photoluminescence measurements were performed with a time-resolved fluorescence microscope (MicroTime 200, PicoQuant GmbH). A detailed description may be found elsewhere.[30] The excitation source is a pulsed diode laser emitting at 482 nm with a repetition rate of 2.5 MHz. Illumination and collection of light are done through a water immersion objective 60x magnification with N.A. of 1.2 (UPLSAPO 60XW, Olympus). Samples are scanned with a piezo xy-stage. The emitted fluorescence is spectrally cleaned through a dichroic mirror and a bandpass filter with transmission in the interval 550–690 nm. A pinhole of 30 μm is used to reject out-of-focus light. The emitted light is detected with single-photon counting avalanche diodes (Perkin-Elmer) and digitized by TimeHarp 200 TC-SPC PC board (PicoQuant GmbH). The instrument response function has a fwhm around 1 ns, and the time increment is 150 ps/channel.

Typically, the image scans performed cover an area of 80×80 μm$^2$ and are composed of (256×256) pixels. The integration time per pixel of 1 ms was low enough to prevent photoenhancement effects in QD emission.[31,32] The excitation power was kept at 108 W/cm$^2$ to minimize multiexcitonic processes, which in these conditions should not exceed 3.1% of excitation probability.[33,34] The luminescence decay integrated over the imaged area can be retrieved from the photon arrival time histogram of all detected photons. The total number of detected photons per image amounted to a few million counts, which afforded reliable decay curves. On the other hand, a direct analysis of the lifetime images is not suitable because only a few counts are detected per pixel and the decays are intrinsically complex. Data analysis with a multi-exponential function was performed with the SymPhoTime software (PicoQuant GmbH). The quality of the fits was evaluated by the usual criteria for the $\chi^2$ parameter and weighted residuals. Decay analysis with Equation 1 (see text for further details) was implemented on Matlab software.

*Results and Discussion*

In order to thoroughly characterize the photoluminescence of QD/polymer films, multiple fluorescence lifetime images were acquired on each film. Different macroscopic positions were scanned to ensure a comprehensive surface sampling. From a total of 25 images acquired per sample, a few selected examples are shown in Figure 1. A similar pattern of surface luminescence was found within each sample. This pattern shows micrometric domains with approximately the same size and emission intensity that cover most of the surface. Some dark areas corresponding to regions not covered are also observed (top right corner of Figure 1A).[8]

Overall, a regular deposition of luminescent material is found on the films prepared by LB

technique. The intensity and lifetime of surface luminescence, however, vary with the deposition method, bilayer or co-spreading deposition, as it is discussed next.

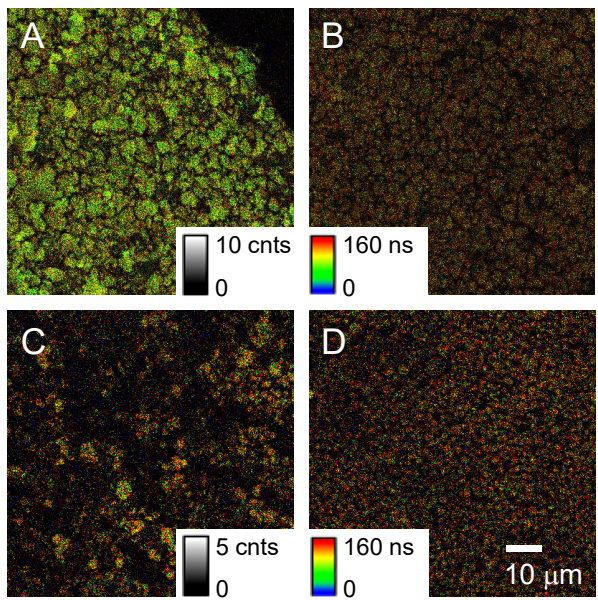

**Figure 1** – Examples of fluorescence lifetime images for CdSe QDs in PSMABEE polymer mixed films: (A) bilayer deposition and (B) co-spreading with a QD-polymer ratio of 1:1, both at high pressure, $\pi$ = 30 mN/m; and co-spreading with 1:2 ratio, (C) at high pressure, $\pi$ = 30 mN/m, and (D) at low pressure, $\pi$ = 14 mN/m. Excitation was selected at 482 nm and emission was collected in the interval 550–690 nm. More examples are shown in section 1 of Supporting Information.

From each image, the luminescence decay is obtained giving a total of 25 decay curves per sample. Examples of luminescence decay curves are shown in Figures 2A and 2B for different conditions of film deposition. Within each sample, the decay curves do not vary much across the sample's surface, as it is illustrated in Figures 2C and 2D, which show the average decay (solid curves) and the area corresponding to the standard deviation multiplied by two (shaded area) from the total of 25 images. The decay curves show a pronounced decrease in the initial hundred ns followed by an exponential tail at long times. A total of four exponential components are required to properly fit these decays. The complete set of decay times fitted by multi-exponential analysis is given in section 2 of Supporting Information. In general, the photoluminescence dynamics of QDs in the ns–µs timescale are intrinsically complex due to the role of surface defect and charge trapped states in excited-state relaxation.[35-39] In the solid films, the complexity is further increased by the possibility of excitation energy transfer between QDs, because of the short interparticle distances in the close-packed film regions.[10-12]

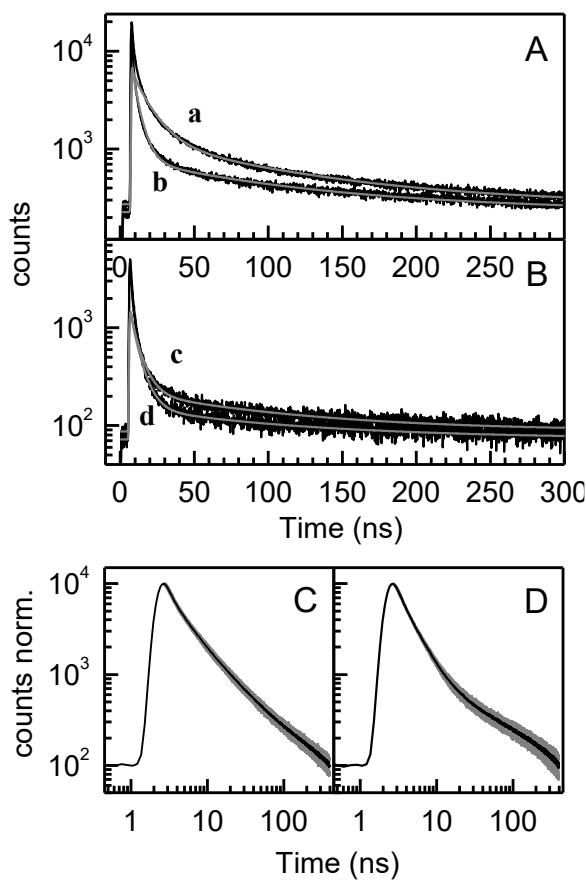

**Figure 2** – Photoluminescence decays of CdSe QDs in PSMABEE polymer mixed films: (A) bilayer deposition (curve a) and co-spreading with a QD-polymer ratio of 1:1 (curve b), both at high pressure, π = 30 mN/m; and (B) co-spreading with 1:2 ratio at high pressure, π = 30 mN/m (curve c) and at low pressure, π = 14 mN/m (curve d). Excitation was at wavelength 482 nm and emission was collected in the interval 550 – 690 nm. Solid curves in grey color show the fitting results with Equation 1 – see text for further details. Normalized and averaged decays of bilayer (C) and co-spreading (D) films over a total of 25 images acquired. The grey shaded area represents the standard deviation multiplied by two.

Prior to the decay analysis, we describe here the qualitative interpretation of the film decays, which is based on the vast literature about photoluminescence dynamics of QDs in solution and deposited in solid films. It has been demonstrated that the detachment of capping molecules can lead to the appearance of short decay components due to quenching by surface defects.[40] Accordingly, we attribute the sharp decrease in the initial nanoseconds to QDs with extensive surface defects that strongly quench emission from band-edge exciton recombination ('$k_{ed}$' in Figure 3).

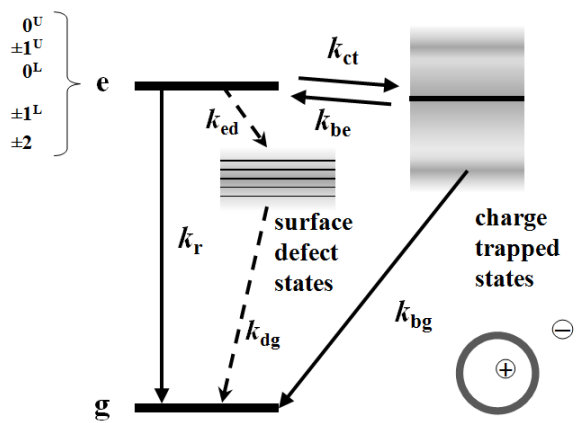

**Figure 3** – Simplified diagram for excited-state processes in QDs. The dashed arrows correspond to additional relaxation pathways from surface-defect sites.

On the other hand, there is evidence that non-exponential photoluminescence decays of QDs are related not only to intraband relaxation, but as well to photoinduced charging processes. In Figure 3, we separate these two processes by distinguishing, respectively, between "surface defect states" and "charge trapped states". However, this is only a formal distinction, as "surface defect states" may also have some character of charge separation and trapping. The key points that we would like to illustrate in this scheme are: *i*) "surface defect states" can be emissive and their spectrum is usually observed at lower energies relatively to core exciton recombination; *ii*) they contribute to the fast deactivation of core exciton states, which shows up as short decay components in their emission. We distinguish these from "charge trapped states", in which case the injected charge is trapped in the surrounding medium of the QD particle but it can recombine to the core excited-state at later times leading to delayed emission, which shows up as long decay components usually with power-law decay kinetics.[37]

Because the emission from surface defect states is usually observed at lower energies relatively to core exciton recombination, the emission spectrum of the QDs dissolved in chloroform and deposited onto quartz were recorded. The emission spectra show the core recombination emission band, centered at 580 nm, and a broad non-structured band centered at wavelengths longer than 700 nm that is attributed to surface defect states (Figures S2 and S3 in the Supporting Information). It is interesting to note that this band is much more pronounced in QDs deposited in films than in chloroform solution. This fact shows that surface defect states are favored by the QDs' deposition. To separate the emission of core exciton and surface defect states, we have recorded the emission decays collected with several filters. The photoluminescence decays practically do not change in the range corresponding to emission from the core exciton recombination band, except for some minor changes that may be attributed to sample heterogeneity. For instance, the decays obtained with filters between 528 and 690 nm are practically superimposable (see more details in section 2 of the Supporting Information). On the other hand, the decays obtained in the wavelength range of 668 – 723 nm, which correspond mostly to surface defects' emission, show a larger contribution from long decay components, as compared to the emission from the core exciton recombination band. Furthermore, the decays obtained for the QDs in solution show an exponential decay tail for the emission from surface defects, while the decay tail from the core exciton recombination band is non-exponential, and is better described by power-law kinetics (Figure S13B in the Supporting Information). All these evidences allow us to assign the longest component in the emission from core exciton recombination to delayed luminescence from charge trapped states. In the remaining of this paper, we will focus our discussion on the effects of film morphology in the photoluminescence decays from core exciton recombination selected with appropriate filters to exclude emission from surface defects. We will also neglect the initial part of the decay curves, t < 10 ns, which corresponds to the fast deactivation processes from intraband relaxation processes by surface defects, and we will focus our data analysis on the long components of emission from core exciton recombination.

In the time range from about ten ns to a hundred ns, the decays exhibit a pronounced curvature that is clearly distinct for films prepared by bilayer and co-spreading deposition (see Figures 2A and B). For comparison in Figure 4A, the results from the multi-exponential analysis of film decays are presented in a simplified plot. In this plot, the short decay times were averaged out using the respective intensity-weights and are represented together with the long decay time. The results obtained for the bilayer film stand out, because the averaged short decay times are clearly longer than for co-spreading films, while the long decay time is quite similar for all samples prepared by the two deposition methods.

The relative brightness of the photoluminescence images also reflects the difference between bilayer and co-spreading films (Figure 4B). Under the same excitation conditions, the bilayer films have a significantly higher intensity of emission compared to co-spreading films, i.e. a larger number of counts per image. This observation could be simply due to a higher surface concentration of QDs in the bilayer films. However, this is not the case, as it is shown in the inset of Figure 4B, where the emission intensity is plotted against the particle density calculated from the surface concentration (see more details in section 4 of Supporting Information). This plot shows that the PL intensity of the co-spreading films is proportional to particle density, while the PL intensity of the bilayer film stands out, indicating that it is much more emissive than the co-spreading films with the same particle density.

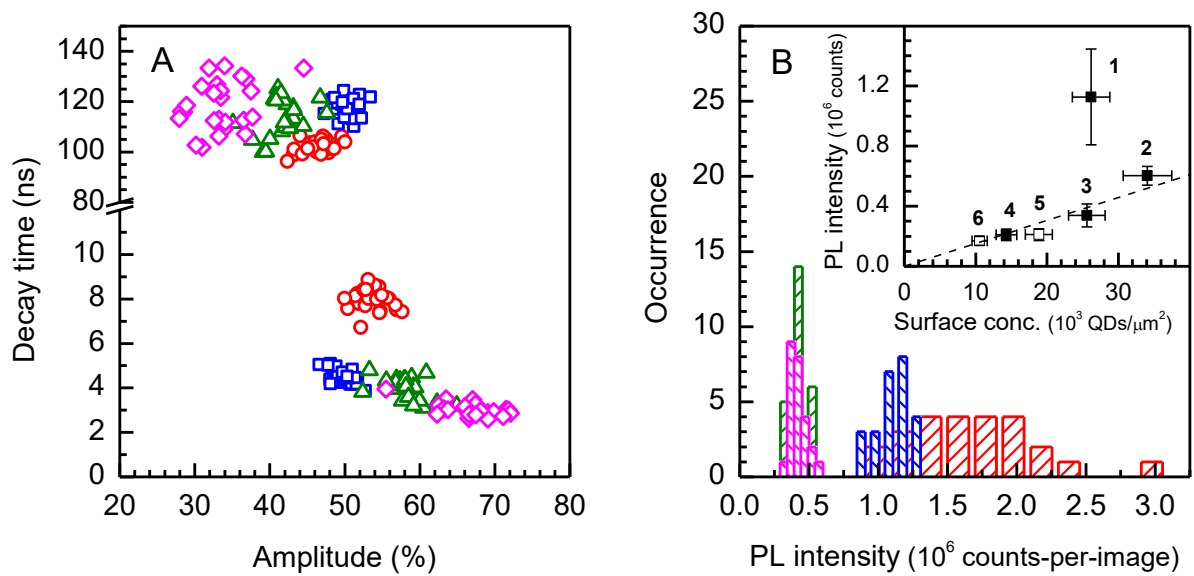

**Figure 4** – (A) Decay times obtained from the multiexponential analysis of each image-integrated decay: the three shorter decay times were averaged out (bottom) and are represented together with the long decay time (top). Results shown are for bilayer deposition (circles) and co-spreading with 1:1 ratio (squares), both at high pressure, π = 30 mN/m; and for co-spreading with 1:2 ratio at high pressure, π = 30 mN/m (triangles), and at low pressure, π = 14 mN/m (diamonds). (B) Photoluminescence (PL) intensity indicated as the total number of counts per image obtained for CdSe QDs in PSMABEE polymer mixed films for bilayer deposition (red), co-spreading with a QD-polymer ratio of 1:1 (blue) at high pressure, π = 30 mN/m; and co-spreading with 1:2 ratio at high pressure, π = 30 mN/m (green) and at low pressure, π = 14 mN/m (magenta). The inset shows the average PL intensity plotted against the surface concentration of QDs for the bilayer (point 1) and co-spreading (points 2-6) films. Closed and open symbols represent data from films deposited at π = 30 mN/m and 14 mN/m, respectively.

In an attempt to interpret the differences between the emission properties of co-spreading and bilayer films, we have analyzed the TEM images of these films. Representative TEM images are shown in Figure 5 and a more detailed analysis is presented in section 4 of the Supporting Information. The TEM images show regions densely covered along with other regions more sparsely covered. Overall, the bilayer film shows a more regular particle distribution than co-spreading films. We have highlighted these regions of more regular packing in the TEM images of bilayer film by a yellow outline in Figure S10 of the Supporting Information. These regions are referred to as "close-packed". The co-spreading films also show close-packed regions, but the packing of dots does not seem so regular, and the presence of small aggregated clusters is more noticeable. We have also highlighted the small aggregated clusters in the TEM images of co-spreading 1:1 film by a red outline in Figure S10. We attribute to these small aggregated clusters of QDs the role of excitation energy traps in the deposited films. This idea is supported by results from other authors showing that luminescence quenching and enhanced blinking occurs in small QD clusters. In order to explain this observation, it was hypothesized that non-emissive charged states could be stabilized by Coulomb interactions between clustered QDs.[41-45]

On the other hand, within an ensemble of QDs, the surface conditions vary between particles, and most likely there is a statistical distribution of the number of surface defects per particle. Some of the QDs are likely to have an extensive number of surface defects, and these will have a low emission intensity and a short decay. We hypothesize that QDs with extensive surface defects may also act as energy traps in our films, in addition to the small aggregated QD clusters, as mentioned to above.

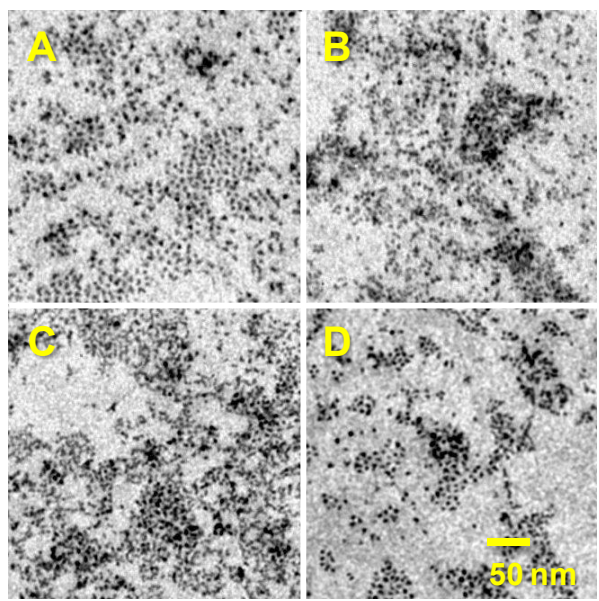

**Figure 5** – Transmission electron microscope images of CdSe QDs in PSMABEE polymer mixed films: (A) bilayer deposition and (B) co-spreading with a QD-polymer ratio of 1:1, both at high pressure, $\pi = 30$ mN/m; and co-spreading with 1:2 ratio, (C) at high pressure, $\pi = 30$

mN/m and (D) at low pressure, π = 14 mN/m. Notice that the patterns of TEM and FLIM images do not find correspondence to each other because of the different length scales (see also section 5 of the Supporting Information).

The photoluminescence dynamics of the films are affected by the surface distribution of particles, because in the regions densely covered the interparticle distances are only a few nm and dot-dot excitation energy transfer is likely to occur within the ensemble of close-packed particles. The migration of excitation energy affects the observed decay because of sample heterogeneity and, in particular, because of excitation energy traps commonly present in solid films, i.e. small aggregated QD clusters or QDs with extensive surface defects.

From about a hundred ns onward, the decays show an almost exponential behavior that is attributed to isolated QDs or to regions of QDs without energy traps. The photoluminescence of QDs decays very slowly with lifetime components longer than hundred ns, and even reaching µs in chloroform solution (see Figure S13B in Supporting Information). This long decay tail observed in the photoluminescence of QDs has been attributed in the literature to a process of delayed luminescence from charge transfer and trapping in the surrounding environment of the quantum-dot.[37,38] Furthermore, it has also been associated with the power-law statistics of emission intermittency of single-dots observed in the timescale of ms to s.[37,46,47] According to this, we assume that delayed luminescence from charge transfer and trapping in the surrounding environment is responsible for the long decay tail observed in the emission of our QDs ('$k_{ct}$' and '$k_{be}$' in Figure 3).

All these considerations allow us to adopt an alternative model to the multi-exponential analysis of QD film decays. In particular, we assume that excitation energy transport and trapping processes in the densely covered regions are the processes most likely competing with radiative emission from core exciton states. The model considers two separate contributions from sparse and compact regions of immobilized QDs in the films. This picture is supported by the TEM images that show regions compactly covered with QDs along with other regions more sparsely covered. Therefore, the decay function assumes two terms, $S(t)$ and $C(t)$, that account separately for sparse and compact regions of quantum-dots,

$$D(t) = \sigma\, S(t) + (1-\sigma)\, C(t). \qquad (1)$$

The $S(t)$ term describes the intrinsic photophysics of QDs, which can be very complex because core exciton states often interact with surface localized states and the surrounding environment. In organically coated QDs, this type of interaction can result in charge trapping on surface defect sites, or on ligand-based orbitals, with the formation of an ionized core state that is usually non-emissive. On the other hand, the term $C(t)$ further accounts for interparticle energy migration and trapping processes, because it describes the contribution from compact film regions. An overview on the models selected for $S(t)$ and $C(t)$, and their respective parameters, is explained below.

To describe the term $S(t)$, we use the model proposed by Tachiya-Mozumder to interpret the luminescence decays affected by recombination from charge trapped to core excited states.[47,48] This model considers a mechanism of charge trapping by electron tunneling toward a uniform distribution of traps. The decay function is calculated by numerical Laplace inversion of,

$$\hat{S}(s) = \frac{1}{s + k_\mathrm{r} + k_\mathrm{ct}\,(\tau_0 s)^\mu}, \qquad (2)$$

with three adjustable parameters, the radiative recombination rate $k_\mathrm{r}$, the charge transfer rate term $\tau_0^\mu k_{ct}$, and a power-law exponent $1 + \mu$. A value of 30 ns was taken as initial guess for $k_r^{-1}$ in the fittings. This value was estimated from the $k_r^{-1}$ fitted for decays of QDs dissolved in chloroform solution and corrected for the dielectric screening in the films, as explained in section 6 of Supporting Information. The charge transfer rate term $\tau_0^\mu k_{ct}$ was varied to fit the long time behavior of the film decays. The initial guess for the power-law exponent was also based on our results of luminescence decays of QDs in chloroform, which afforded $1 + \mu \sim 2$, or equivalently $\mu \sim 1$. In this case, from Equation 2 we have,

$$S(t) \approx \exp[-k_r/(1 + k_{ct}\tau_0) \times t]. \qquad (3)$$

Here, we notice that Eq. 3 predicts a decay law exponential, in agreement with results observed at long times in the decays of Figure 2.

To describe the term related with the energy transport and trapping, term $C(t)$ in Equation 1, we adopt the model developed by Fayer and co-workers.[49,50] This model was proposed to interpret the excitation transfer in disordered two dimensional systems with randomly distributed donor and trap species. In the model, the transport master equation uses the diagrammatic expansion of the Green's function to obtain the probability of finding an excitation at a given position $r$ at time $t$. Accordingly, the Green's function $G^s(r - r', t)$ for the probability that excitation is on the initially excited site at time $t$ is given by the following self-consistent equation in Laplace domain,

$$\hat{G}^s(\epsilon) + \left(2^{1/3}\pi/3^{3/2}\tau_D^{1/3}\epsilon\right)\left(C_D + 2^{2/3}C_T\right)[\hat{G}^s(\epsilon)]^{1/3} - 1/\epsilon = 0, \qquad (4)$$

with $\tau_D$, the donor lifetime, $C_D$ and $C_T$, the donor and trap reduced concentrations, as described below. From Equation 4 the probability $P_D(t)$ that an excitation is in the donor ensemble at time $t$ is obtained from inverse transform of,

$$\hat{G}^D(0,\epsilon) = \hat{G}^s(\epsilon)/\left\{1 - \left(2^{1/3}\pi\, C_D/3^{3/2}\tau_D^{1/3}\right)[\hat{G}^s(\epsilon)]^{1/3}\right\}. \tag{5}$$

The corresponding decay law is obtained multiplying $P_D(t)$ by a decaying exponential with the donor lifetime,

$$C(t) = \exp(-t/\tau_D) \times P_D(t). \tag{6}$$

When fitting the decays, we have assumed the approximation in Equation 3 for the individual QD decay in the timescale of hundreds of ns, which implies that the donor lifetime is $\tau_D \sim k_r^{-1}(1 + k_{ct}\tau_0)$. The reduced concentrations $C_D$ and $C_T$ are defined by,

$$C_D = \pi(R^{DD})^2 \rho_D \quad , \quad C_T = \pi(R^{DT})^2 \rho_T, \tag{7}$$

with $R^{DD}$ and $R^{DT}$, the Förster radius for donor-donor and donor-trap energy transfer, $\rho_D$ and $\rho_T$, the number densities of donors and traps, respectively. Although the interparticle distance is comparable with the QD size, it has been shown that the dipolar approximation in Förster theory describes reasonably well the excitation energy transfer between QDs or even between QD and fluorescent molecules.[51,52] Using the theory, we estimate a Förster radius $R^{DD}$ of 4.1 nm for donor-donor energy transfer and assume a similar $R^{DT}$ value for donor-trap energy transfer. However, when fitting the decays no assumption was made about $R^{DD}$ and $R^{DT}$ values by treating $C_D$ and $C_T$ as the adjustable parameters.

The decays of our QD films are only approximately described by Equation 1 mainly because of two reasons: *i*) at very short times (below ten ns), the surface defect states contribute predominantly to fast excited-state decay; and *ii*) the model considered for the compact regions, $C(t)$, does not account for boundary effects due to finite domain size. In spite of these limitations, the approximate description of Equation 1 allows to rationalize the photoluminescence dynamics of QD films using only a few adjustable parameters that afford physical insight about the systems studied.

The fittings with Equation 1 to the decays of bilayer and co-spreading films are shown in Figures 2A and 2B (curves a – d) and the parameters obtained are given in Table 1. The radiative lifetime $k_r^{-1}$ of ca. 30 ns is similar for every film, as expected, since it is an intrinsic photophysical property of the QDs. It agrees with the typical values of radiative lifetime of CdSe QDs reported in the literature.[4] The long decay components, in the range of hundreds of ns (see Figure 4), correspond to luminescence decay times that are affected by other non-radiative processes, such as photoinduced charge separation and trapping, which by recombination to the core excited-states leads to delayed emission. The spectral distribution of the long decay components was shown to essentially follow the emission spectrum of the core exciton recombination band,[53] with slightly more contribution in the low energy side but that nonetheless strongly supports that it corresponds to emission from core excited-states. The model defined by Equation 1 explicitly describes the process of delayed emission through the parameters $\tau_0^\mu k_{ct}$, and $1 + \mu$, as described before.

On the other hand, it is interesting to notice that the concentration of donors, $C_D$, obtained from the fit process is almost the same for both systems and it is close to the value of 3.3 particles within the area of a Förster radius, while the concentration of traps, $C_T$, is approximately double for bilayers than for co-spreading films. The dense concentration of donors $C_D$ is consistent with the information obtained from TEM. Thus, if one considers the estimated Förster radius of 4.1 nm and the QD average size of 3.5 nm, then a hexagonal compact arrangement would be achieved for a interparticle surface-to-surface distance of about 0.7 nm, which compares well with the length of an octylphosphonate derivative, e.g. the TOPO molecule.[54] This is in agreement with the simple visual inspection of TEM images of bilayer films, which show compact regions with an approximately geometrical arrangement of particles (see Figure 4, or Figures S5 and S10 in the Supporting Information). However, to clarify this issue we have calculated the radial distribution function (RDF) from the distribution of interparticle center distances in TEM images. This function provides information about the short-range order inside the QDs domains. The RDF of the TEM images from the bilayer film shows a profile with two peaks positioned at 5.1 ± 0.7 and 9.8 ± 0.7 nm. These features compare roughly with the peak positions for the first two shells of a hexagonal compact arrangement with the particle size and separation referred to above (see Figure S11 in Supporting Information). By contrast, the RDF of co-spreading films shows a single peak at 6.2 ± 0.7 nm followed by a plateau. The differences in the radial distribution function give support to a more regular structure in the bilayer film, as previously inferred from a visual inspection of the TEM images, and further reinforce the role of film morphology in the photoluminescence properties of QD films.

**Table 1** – Parameters fitted with Equation 1 to photoluminescence decays of CdSe QDs in PSMABEE polymer mixed films for the several deposition conditions – see text for further details.

|  | High π | | | | Low π | |
| --- | --- | --- | --- | --- | --- | --- |
|  | **Bilayer** | **Co-spr (1:1)** | **Co-spr (1:2)** | **Co-spr (2:1)** | **Co-spr (1:2)** | **Co-spr (2:1)** |
| $k_r^{-1}$ (ns) | 30.0 | 30.0 | 30.1 | 30.0 | 30.0 | 30.0 |
| $\tau_0^\mu k_{ct}$ | 1.99 | 1.92 | 1.99 | 1.97 | 2.00 | 2.01 |
| $1 + \mu$ | 1.98 | 1.92 | 1.96 | 1.96 | 1.92 | 1.97 |
| $\sigma$ | 0.345 | 0.172 | 0.195 | 0.197 | 0.136 | 0.074 |
| $C_D$ | 3.30 | 3.34 | 3.34 | 3.28 | 3.34 | 3.29 |
| $C_T$ | 0.281 | 0.670 | 0.536 | 0.443 | 0.566 | 0.640 |

The decays from both bilayer and co-spreading films are approximately exponential at long times with the same slope (Figure 2A,B). Thus, the model parameters $k_r^{-1}$, $\tau_0^\mu k_{ct}$, and $1+\mu$ describing the decay of isolated QDs or domains without energy traps, $S(t)$, are similar for bilayer and co-spreading films. Only the relative contribution to the decay, $\sigma$, is different with the bilayer film having a larger contribution of 34.5 % compared to 17.2 % in the co-spreading film with QD-polymer 1:1 ratio. We relate the larger contribution of $S(t)$ in bilayer deposition to a film structure that contains a lower number of small aggregated QD clusters that act as energy traps. The cluster formation has been previously studied and was attributed to dewetting processes. In our systems, previous results have demonstrated that the deposition of QDs on LB films of polymers or Gemini surfactants avoids the dewetting processes produced by co-spreading deposition.[15,16] The current work further supports that nanoparticle clustering acts as energy traps thereby reducing the PL emission, as it was proposed elsewhere.[44]

To evaluate the effect of the polymer concentration on the luminescence properties, the photoluminescence decays of films with different QD-to-polymer molar ratio were analyzed according to Equation 1. The QD-polymer molar ratios selected were 1:1, 1:2 and 2:1 respectively. The parameter values obtained from the decay fits show no significant differences, except for a small increase of the $S(t)$ contribution when the polymer concentration increases. More significant is the emission brightness observed in these films that is related to the distinct surface density of QDs, as it is shown in the inset of Figure 4B.

The effect of surface pressure in co-spreading deposition was also evaluated. The fitted parameters obtained for decays curves of films prepared by co-spreading method and surface pressure of 14 mN/m are collected in Table 1. From the parameter values it is possible to conclude that at low surface pressure the film decays have a lower contribution from the long time exponential component (Figure 2B), while the rest of fitted parameters are almost independent of the polymer concentration. The decrease of the emission intensity found in these films is also consistent with the decrease of the QD surface concentration (see inset of Figure 4B).

*Conclusions*

The photoluminescence of QD/polymer mixed films was characterized for different conditions of film deposition: bilayer or co-spreading; QD-to-polymer composition; and surface pressure at deposition. Results show that the photoluminescence dynamics of QD films are affected by energy transport and trapping processes. The efficiency of these processes depends essentially on inter-dot distance and on surface density of energy traps, which can be related to QD clusters or QDs with extensive surface defects. Among the several conditions studied, the bilayer deposition yielded the films with more emission intensity, i.e. more brightness, than those built by co-spreading deposition. The simultaneous analysis of photoluminescence decays and TEM images allowed us to relate this behavior to a lower amount of QD clustering in the bilayer film. We propose a model to interpret the photoluminescence decays that considers energy transport and trapping in the films and, although it is only an approximate model, it also allowed us to relate the photoluminescence dynamics to film morphology. Indeed, the concentration of energy traps retrieved from decay analysis for the bilayer film is about half of those in co-spreading films. Finally, from our results it is possible to conclude that to improve the photoluminescence properties of QD films, it is paramount to minimize the surface defects and the QD clustering, which cause pronounced emission quenching.

**Supporting Information.** Photoluminescence images of QD/polymer films; Results from multi-exponential decay analysis; Photoluminescence decays of CdSe QDs in solution and polymer films spectrally filtered; Analysis of TEM images of QD/polymer films; Comparison of image sizes from optical microscopy and TEM; Photoluminescence dynamics of CdSe QDs in solution. This material is available free of charge via the Internet at http://pubs.acs.org.

**Corresponding author e-mail:** pedro.m.paulo@ist.utl.pt

*Acknowledgements*

The authors thank financial support from ERDF and MEC (MAT 2010-19727). B.M.G. wishes to thank the European Social Fund and Consejería de Educación de la Junta de Castilla y León for her FPI grant and the research fellowship for the stay at the Molecular Photochemistry Group (CQE/IST). We also thank to Microscopy Electron Service (Universidad de Salamanca) for the TEM measurements. P.M.R.P. acknowledges Program Ciência 2008 from FCT. Thanks are due to Fundação para a Ciência e a Tecnologia (FCT, Portugal) for 3° Quadro Comunitário de Apoio (FEDER); FCT/Re-equipment Project 115/QUI/2005.

*References and notes*

1. Coe, S.; Woo, W-K.; Bawendi, M.; Bulovic, V. Electroluminescence from single monolayers of nanocrystals in molecular organic devices. *Nature* **2002**, *420*, 800-803.

2. Talapin, D. V.; Lee, J.-S.; Kovalenko, M. V.; Shevchenko, E. V. Prospects of Colloidal Nanocrystals for Electronic and Optoelectronic Applications. *Chem. Rev.* **2010**, *110*, 389-458.

3. Saunders, B. R. Hybrid polymer/nanoparticle solar cells: Preparation, principles and challenges. *J. Colloid Interface Sci.* **2012**, *369*, 1-15.

4. *Semiconductor Nanocrystal Quantum Dots: Synthesis, Assembly, Spectroscopy and Applications*, Rogach, A. L. (Ed.); Springer: Wien – New York, 2008, pag. 277–310 and references therein.

5. Tomczak, N.; Janczewski, D.; Han, M.; Vancso, G. J. Designer polymer-quantum dot architectures. *Prog. Polym. Sci.* **2009**, *34*, 393-430.

6. Kaufmann, S.; Stöferle, T.; Moll, N.; Mahrt, R. F.; Scherf, U.; Tsami, A.; Talapin, C. V.; Murray, C. B. Resonant energy transfer within a colloidal nanocrystal polymer host system. *Appl. Phys. Lett.* **2007**, *90*, 071108.

7. Chen, C. W.; Wang, C. H.; Chen, Y. F.; Lai, C. W.; Chou, P. T. Tunable energy transfer efficiency based on the composite of mixed CdSe quantum dots and elastomeric film. *Appl. Phys. Lett.* **2008**, *92*, 051906.

8. Issac, A.; von Borczyskowski, C.; Cichos, F. Correlation between photoluminescence intermittency of CdSe quantum dots and self-trapped states in dielectric media. *Phys. Rev. B* **2005**, *71*, 161302(R).

9. Issac, A.; Krasselt, C.; Cichos, F.; von Borczyskowski, C. Influence of the Dielectric Environment on the Photoluminescence Intermittency of CdSe Quantum Dots. *ChemPhysChem* **2012**, *13*, 3223-3230.

10. Kagan, A. R.; Murray, C. B.; Bawendi, M. G. Long-range resonance transfer of electronic excitations in close-packed CdSe quantum-dot solids. *Phys. Rev. B* **1996**, *54*, 8633–8643.

11. Achermann, M.; Petruska, M. A.; Crooker, S. A.; Klimov, V. I. Picosecond Energy Transfer in Quantum Dot Langmuir−Blodgett Nanoassemblies. *J. Phys. Chem. B* **2003**, *107*, 13782-13787.

12. Lunz, M.; Bradley, A. L.; Gerard, V. A.; Byrne, S. J.; Gun'ko, Y. K.; Lesnyak, V. ; Gaponik, N. Concentration dependence of Förster resonant energy transfer between donor and acceptor nanocrystal quantum dot layers: Effect of donor-donor interactions. *Phys. Rev. B* **2011**, *83*, 115423.

13. Tao, A. R.; Huang, J.; Yang, P. Langmuir−Blodgettry of Nanocrystals and Nanowires. *Acc. Chem. Res.* **2008**, *41*, 1662-1673.

14. Park, J. Y.; Advincula, R. C. Nanostructuring polymers, colloids, and nanomaterials at the air–water interface through Langmuir and Langmuir–Blodgett techniques. *Soft Matter* **2011**, *7*, 9829-9843.

15. Alejo, T.; Merchán, M. D.; Velázquez, M. M.; Pérez-Hernández, J. A. Polymer/surfactant assisted self-assembly of nanoparticles into Langmuir-Blodgett films. *Mater. Chem. Phys.*

**2013**, *138*, 286-294.

16.     Martín-García, B.; Velázquez, M.M. Block copolymer assisted self-assembly of nanoparticles into Langmuir-Blodgett films: Effect of polymer concentration. *Mater. Chem. Phys.* **2013** (http://dx.doi.org/10.1016/j.matchemphys.2013.05.017)

17.     Jones, R.; Winter, C. S.; Tredgold, R. H.; Hodge, P.; Hoorfar, A. Electron-beam resists from Langmuir-Blodgett films of poly(styrene/maleic anhydride) derivatives. *Polymer* **1987**, *28*, 1619-1626.

18.     Collins, S. J.; Mary, N. L.; Radhakrishnan, G.; Dhathathreyan, A. Studies of spread monolayers of derivative of styrene–maleic anhydride copolymers. *J. Chem. Soc. Faraday Trans.* **1997**, *93*, 4021-4023.

19.     Jones, M.; Scholes, G. D. On the use of time-resolved photoluminescence as a probe of nanocrystal photoexcitation dynamics. *J. Mater. Chem.* **2010**, *20*, 3533–3538.

20.     Yu, W. W.; Peng, X. Formation of High-Quality CdS and Other II–VI Semiconductor Nanocrystals in Noncoordinating Solvents: Tunable Reactivity of Monomers. *Angew. Chem. Int. Ed. Engl.* **2002**, *41*, 2368-2371.

21.     Kopping, J. T.; Patten, T. E. Identification of Acidic Phosphorus-Containing Ligands Involved in the Surface Chemistry of CdSe Nanoparticles Prepared in Tri-N-octylphosphine Oxide Solvents. *J. Am. Chem. Soc.* **2008**, *130*, 5689-5698.

22.     Owen, J. S.; Park, J.; Trudeau, P.-E.; Alivisatos, A. P. Reaction Chemistry and Ligand Exchange at Cadmium−Selenide Nanocrystal Surfaces. *J. Am. Chem. Soc.* **2008**, *130*,

12279-12281.

23. Morris-Cohen, A. J.; Donakowski, M. D.; Knowles, K. E.; Weiss E. A. The Effect of a Common Purification Procedure on the Chemical Composition of the Surfaces of CdSe Quantum Dots Synthesized with Trioctylphosphine Oxide. *J. Phys. Chem. C* **2010**, *114*, 897–906.

24. Kern, W. The Evolution of Silicon Wafer Cleaning Technology. *J. Electrochem. Soc.* **1990**, *137*, 1887-1892.

25. Gattás-Asfura, K. M.; Constantine, C. A.; Lynn, M. J.; Thimann, D. A.; Ji, X.; Leblanc, R. M. Characterization and 2D Self-Assembly of CdSe Quantum Dots at the Air−Water Interface. *J. Am. Chem. Soc.* **2005**, *127*, 14640-14646.

26. Martín-García, B.; Velázquez, M. M.; Pérez-Hernández, J. A.; Hernández-Toro, J. Langmuir and Langmuir−Blodgett Films of a Maleic Anhydride Derivative: Effect of Subphase Divalent Cations. *Langmuir* **2010**, *26*, 14556-14562.

27. Kalyuzhny, G.; Murray, R. W. Ligand Effects on Optical Properties of CdSe Nanocrystals. *J. Phys. Chem. B* **2005**, *109*, 7012-7021.

28. Bullen, C.; Mulvaney, P. The Effects of Chemisorption on the Luminescence of CdSe Quantum Dots. *Langmuir* **2006**, *22*, 3007-3013.

29. Munro, A. M.; Plante, I. J.-L.; Ng, M. S.; Ginger, D. S. Quantitative Study of the Effects of Surface Ligand Concentration on CdSe Nanocrystal Photoluminescence. *J. Phys. Chem. C* **2007**, *111*, 6220-6227.

30. Paulo, P. M. R.; Costa, S. M. B. Single-Molecule Fluorescence of a Phthalocyanine in PAMAM Dendrimers Reveals Intensity−Lifetime Fluctuations from Quenching Dynamics. *J. Phys. Chem. C* **2010**, *114*, 19035–19043.

31. Šimurda, M.; Němec, P.; Trojánek, F.; Malý, P. Substantial enhancement of photoluminescence in CdSe nanocrystals by femtosecond pulse illumination. *Thin Solid Films* **2004**, *453–454*, 300–303.

32. Pechstedt, K.; Whittle, T.; Baumberg, J.; Melvin, T. Photoluminescence of Colloidal CdSe/ZnS Quantum Dots: The Critical Effect of Water Molecules. *J. Phys. Chem. C* **2010**, *114*, 12069–12077.

33. Fisher, B.; Caruge, J. M.; Zehnder, D.; Bawendi, M. Room-Temperature Ordered Photon Emission from Multiexciton States in Single CdSe Core-Shell Nanocrystals. *Phys. Rev. Lett.* **2005**, *94*, 087403.

34. Fisher, B.; Caruge, J.-M.; Chan, Y.-T.; Halpert, J.; Bawendi, M. G. Multiexciton fluorescence from semiconductor nanocrystals. *Chem. Phys.* **2005**, *318*, 71–81.

35. Bawendi, M. G.; Carroll, P. J.; Wilson, W. L.; Brus, L. E. Luminescence properties of cadmium selenide quantum crystallites: resonance between interior and surface localized states. *J. Chem. Phys.* **1992**, *96*, 946-954.

36. Fisher, B. R.; Eisler, H.-J.; Stott, N. E.; Bawendi, M. G. Emission Intensity Dependence and Single-Exponential Behavior In Single Colloidal Quantum Dot Fluorescence Lifetimes. *J. Phys. Chem. B* **2004**, *108*, 143-148.

37. Sher, P. H.; Smith, J. M.; Dalgarno, P. A.; Warburton, R. J.; Chen, X.; Dobson, P. J.; Daniels, S. M.; Pickett, N. L.; O'Brien, P. Power law carrier dynamics in semiconductor nanocrystals at nanosecond timescales. *Appl. Phys. Lett.* **2008**, *92*, 101111.

38. Jones, M.; Lo, S. S.; Scholes, G. D. Signatures of Exciton Dynamics and Carrier Trapping in the Time-Resolved Photoluminescence of Colloidal CdSe Nanocrystals. *J. Phys. Chem. C* **2009**, *113*, 18632–18642.

39. Nadeau, J. L.; Carlini, L.; Suffern, D.; Ivanova, O.; Bradforth, S. E. Effects of β-Mercaptoethanol on Quantum Dot Emission Evaluated from Photoluminescence Decays. *J. Phys. Chem. C* **2012**, *116*, 2728−2739.

40. Hartmann, L.; Kumar, A.; Welker, M.; Fiore, A.; Julien-Rabant, C.; Gromova, M.; Bardet, M.; Reiss, P.; Baxter, P. N. W.; Chandezon, F.; Pansu, R. B. Quenching Dynamics in CdSe Nanoparticles: Surface-Induced Defects upon Dilution. *ACS Nano* **2012,** *6*, 9033-9041.

41. Koole, R.; Liljeroth, P.; de Mello Donegá, C.; Vanmaekelbergh, D.; Meijerink, A. Electronic Coupling and Exciton Energy Transfer in CdTe Quantum-Dot Molecules. *J. Am. Chem. Soc.* **2006**, *128*, 10436-10441.

42. Yu, M.; Van Orden, A. Enhanced Fluorescence Intermittency of CdSe-ZnS Quantum-Dot Clusters. *Phys. Rev. Lett.* **2006**, *97*, 237402.

43. Lee, J. D.; Maenosono, S. Intensified blinking, continuous memory loss, and fluorescence enhancement of interacting light-emission quantum dots. *Phys. Rev. B* **2009**, *80*, 205327.

44. Shepherd, A. P.; Whitcomb, K. J.; Milligan, K. K.; Goodwin, P. M.; Gelfand, M. P.;

Van Orden, A. Fluorescence Intermittency and Energy Transfer in Small Clusters of Semiconductor Quantum Dots. *J. Phys. Chem. C* **2010**, *114*, 14831–14837.

45. The Coulombic stabilization of charged states only induces quenching if it favors charge recombination to ground state relative to excited state (respectively, '$k_{bg}$' and '$k_{be}$' in Figure 3).

46. Cichos, F.; von Borczyskowski, C.; Orrit, M. Power-law intermittency of single emitters. *Curr. Opin. Colloid Interface Sci.* **2007**, *12*, 272–284.

47. Tachiya, M.; Seki, K. Unified explanation of the fluorescence decay and blinking characteristics of semiconductor nanocrystals. *Appl. Phys. Lett.* **2009**, *94*, 081104.

48. Tachiya, M.; Mozumder, A. Kinetics of geminate-ion recombination by electron tunnelling. *Chem. Phys. Lett.* **1975**, *34*, 77–79.

49. Gochanour, C. R.; Andersen, H. C.; Fayer, M. D. Electronic excited state transport in solution. *J. Chem. Phys.* **1979**, *70*, 4254–4271.

50. Loring, R. F.; Fayer, M. D. Electronic excited state transport and trapping in one-and two-demensional disordered systems. *Chem. Phys.* **1982**, *70*, 139-147.

51. Curutchet, C.; Franceschetti, A.; Zunger, A.; Scholes, G. D. Examining Förster Energy Transfer for Semiconductor Nanocrystalline Quantum Dot Donors and Acceptors. *J. Phys. Chem. C* **2008**, *112*, 13336–13341.

52. Crooker, S. A.; Hollingsworth, J. A.; Tretiak, S.; Klimov, V. I. Spectrally Resolved Dynamics of Energy Transfer in Quantum-Dot Assemblies: Towards Engineered Energy Flows in Artificial Materials. *Phys. Rev. Lett.* **2002**, *89*, 186802.

53. Petrov, E. P.; Cichos, F.; von Borczyskowski, C. Intrisic photophysics of Semiconductor Nanocrystals In Dielectric Media: Formation of Surface States. *J. Luminescence* **2006**, *119-120*, 412-417.

54. Jiang, J.; Krauss, T. D.; Brus, L. E. Electrostatic Force Microscopy Characterization of Trioctylphosphine Oxide Self-assembled Monolayers on Graphite. *J. Phys. Chem. B* **2000**, *104*, 11936-11941.

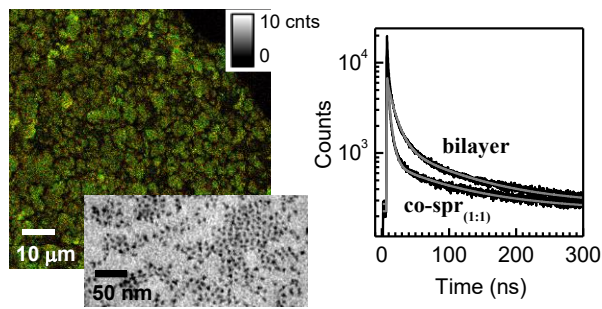

**Graphical TOC.** Photoluminescence Dynamics of CdSe QD/polymer Langmuir-Blodgett Thin Films: Morphology Effects